\documentclass[preprint,showpacs,amsmath,amssymb,aps]{revtex4}
\usepackage{graphicx}
\begin{document}

\title{Semi-phenomenological description of the chiral bands in $^{188,190}Os$}

\author{A. A. Raduta$^{a),b)}$ and  C. M. Raduta$^{a)}$ }

\address{$^{a)}$ Department of Theoretical Physics, Institute of Physics and
  Nuclear Engineering, Bucharest, POBox MG6, Romania}

\address{$^{b)}$Academy of Romanian Scientists, 54 Splaiul Independentei, Bucharest 050094, Romania}

\begin{abstract}
A set of interacting particles are coupled to a phenomenological core described using the generalized coherent state model. Among the particle-core states a finite set which have the property that the angular momenta carried by the proton and neutron quadrupole bosons and the particles, separately, are mutually orthogonal are identified.  The magnetic properties of such  states are studied. All terms of the model Hamiltonian exhibit chiral symmetry except the spin-spin interaction. There are four bands of the type with two-quasiparticle-core dipole states, exhibiting properties which are specific for magnetic twin bands. An application is presented, for the isotopes $^{188, 190}$Os.
\end{abstract}


\maketitle

\renewcommand{\theequation}{1.\arabic{equation}}
\setcounter{equation}{0}
\section{Introduction}

 Some of the fundamental properties of nuclear systems may be evidenced through their  interaction with an electromagnetic field. The two components of the field, electric and magnetic, are used to explore the properties of electric and magnetic nature, respectively. At the end of the last century, the scissors like states \cite{LoIu1,LoIu2,LoIu3} and the spin-flip excitations \cite{Zawischa} were widely treated by various groups. The scissors mode provides a description of the angular oscillation of the proton against a neutron system, and the total strength is proportional to the nuclear deformation squared, which reflects the collective character of the excitation \cite{LoIu3,Zawischa}. 

By virtue of this feature it was believed that the magnetic collective properties are in general associated with deformed systems. This is not true due to the magnetic dipole bands, where the ratio between the moment of inertia and the B(E2) value for exciting the first $2^+$ from the ground state $0^+$,
${\cal I}^{(2)}/B(E2)$,
takes large values, of the order of 100(eb)$^{-2}MeV^{-1}$. These large values can be explained
by there being a large transverse magnetic dipole moment which induces dipole magnetic transitions, but almost no charge quadrupole moment \cite{Frau}. Indeed, there are several experimental data sets showing that the dipole bands have large values for $B(M1)\sim 3-6\mu^2_N$, and very small values of $B(E2)\sim 0.1(eb)^2$ (see for example Ref.\cite{Jenkins}). The states are different from the scissors mode ones, exhibiting instead a shears character. A system with a large transverse magnetic dipole moment may consist of a triaxial core to which a proton prolate  orbital and a  neutron oblate hole orbital are coupled. The maximal transverse dipole momentum is achieved when, for example, $\bf{j}_p$ is oriented along the small axis of the core and $\bf{j}_n$ along the long axis and the core rotates around the intermediate axis.   Suppose that the three orthogonal angular momenta form a right trihedral frame. If the Hamiltonian describing the interacting system of protons, neutrons and the triaxial core is invariant to the transformation which changes the orientation of one of the three angular momenta, i.e. the right trihedral frame is transformed to one of a left type, one says that the system exhibits a chiral symmetry. As always happens, such a symmetry is identified when it is broken and consequently to the two trihedral ones there correspond distinct energies, otherwise close to each other. Thus, a signature for a chiral symmetry characterizing a triaxial system is the existence of two $\Delta I=1$ bands which are close in energies.  On increasing the total angular momentum, the gradual alignment of $\bf{j}_p$ and $\bf{j}_n$ to the total $\bf{J}$ takes place and a magnetic band is developed.

In \cite{AAR2014} we attempted to investigate another chiral system consisting of one phenomenological core with two components, one for protons and one for neutrons, and two quasiparticles whose angular momentum ${\bf J}$ is oriented along the symmetry axis of the core due to the particle-core interaction. In the quoted reference we proved that states of total angular momentum ${\bf I}$, where the three components mentioned above carry the angular momenta ${\bf J_p}, {\bf J_n}, {\bf J}$  which are mutually orthogonal, do exist. Such a configuration seems to be optimal for defining a large transverse magnetic moment that induces large M1 transitions. In choosing the candidate nuclei with chiral features, we were guided by the suggestion \cite{Frau} that triaxial nuclei may favor orthogonality of the aforementioned three angular momenta and therefore may exhibit a large transvercse magnetic moment. In the previous publication, the formalism was applied to  $^{192}$Pt, which satisfies the triaxiality signature condition.

Here the same formalism is applied to two other isotopes, $^{188,190}$Os. Moreover the proton and neutrn gyromagnetic factors are calculated in a self-consistent manner. Also, an extended discussion concerning chiral symmetries of the spin-spin interaction, the broken symmetries and associated phase transition is presented.

\section{Brief review of the GCSM}
\renewcommand{\theequation}{2.\arabic{equation}}
\setcounter{equation}{0}

The core is described by the Generalized Coherent State Model (GCSM)\cite{Rad2} which is an extension of the Coherent State Model (CSM)\cite{Rad1} for a composite system of protons and neutrons. 
The CSM is based of the ingredients presented below.

For the sake of giving a self-contained presentation, in what follows we shall give some minimal information about the phenomenological formalism of GCSM, providng a description of the core system. In his way the necessary  notation and the specific properties of the core are presented.
The usual procedure used for describing the excitation energies with a given boson Hamiltonian is to diagonalize it and fix the structure coefficients such that some particular energy levels are reproduced. For a given angular momentum, the lowest levels belong to the ground, gamma and beta bands, respectively. For example, the lowest  state of angular momentum 2, i.e. $2^+_1$, is a ground band state, the next lowest, $2^+_2$, is a gamma band state, while $2^+_3$ belongs to the $\beta $ band. The dominant components of the corresponding eigenstates are one, two and three phonon states. The harmonic limit of the model Hamiltonian yields a multi-phonon spectrum while on switching on a deforming anharmonicity, the spectrum is a reunion of rotational bands. The correspondence of the two kinds of spectra, characterizing the vibrational and rotational regimes respectively, is realized according to the Sheline-Sakai scheme \cite{Sheline}. In the near vibrational limit a certain staggering is observed for the $\gamma$ band, while in the rotational extreme, the staggering is different. The bands are characterized by the quantum number $K$ which for the axially symmetric nuclei is 0 for the ground and $\beta$ bands and equal to 2 for $\gamma$ band. The specific property of a band structure consists in the E2 probabilities of transition within a band being much larger that the ones for transitions between two different bands. For $\gamma$ stable nuclei, the energies of the states heading the $\gamma$ and $\beta$ bands are ordered as $E_{2^{+}_{\gamma}}> E_{0^{+}_{\beta}}$, while for $\gamma$ unstable nuclei the ordering is reversed. A third class of nuclei  exist for which $E_{J^{+}_{\gamma}}\approx E_{J^{+}_{\beta}}$, $J$-even.
{\it These are the fundamental features for which the wave functions should provide a description in any realistic approach}. The CSM builds a restricted basis requiring that the states
 before and after angular momentum projection are orthogonal and, moreover, accounts for the properties listed above. If such a construction is possible, then one attempts to define an effective Hamiltonian which is quasi-diagonal in the selected basis. The CSM is, as a matter of fact, a possible solution in terms of quadrupole bosons \cite{Rad1}.

Unlike within the CSM,  within the GCSM \cite{Rad2} quadrupole
proton-like bosons, $b^{\dagger}_{p\mu}$, provide the description of the protons, while quadrupole neutron-like bosons, $b^{\dagger}_{n\mu}$, provide that of neutrons.
Since one deals with two quadrupole bosons instead of one, one expects 
to have a more flexible model and to find a simpler solution satisfying the restrictions
required by CSM.  The restricted  collective space is defined  by the states providing the description of the three
major bands: ground, beta and gamma, as well as the band  based on
the  isovector state $1^+$. Orthogonality conditions, required for both intrinsic and projected states, are satisfied by the
following six functions which generate, by angular momentum projection,
six rotational bands:
\begin{eqnarray}
|g;JM\rangle&=&N^{(g)}_JP^J_{M0}\psi_g,~~
|\beta;JM\rangle = N^{(\beta)}_JP^J_{M0}\Omega_{\beta}\psi_g,~~
|\gamma;JM\rangle = N^{(\gamma)}_JP^J_{M2}(b^{\dag}_{n2}-b^{\dag}_{p2})\psi_g,
\nonumber\\
|\tilde{\gamma} ;JM\rangle&=&N^{(\tilde{\gamma})}_JP^J_{M2}(\Omega^{\dag}_{\gamma,p,2}+\Omega^{\dag}_{\gamma,n,2})\psi_g,~~
|1;JM\rangle = N^{(1)}_JP^J_{M1}(b^{\dag}_nb^{\dag}_p)_{11}\psi_g,\nonumber\\
|{\tilde 1};JM\rangle &=& N^{(\tilde{1})}_JP^J_{M1}(b^{\dag}_{n1}-b^{\dag}_{p1})\Omega^{\dag}_{\beta}\psi_g,
\psi_g = exp[(d_pb^{\dag}_{p0}+d_nb^{\dag}_{n0})-(d_pb_{p0}+d_nb_{n0})]|0\rangle .
\label{figcsm}
\end{eqnarray}
Here, the following notations have been used:
\begin{eqnarray}
\Omega^{\dag}_{\gamma,k,2}&=&(b^{\dag}_kb^{\dag}_k)_{22}+d_k\sqrt{\frac{2}{7}}
b^{\dag}_{k2},~~\Omega^{\dag}_k=(b^{\dag}_kb^{\dag}_k)_0-\sqrt{\frac{1}{5}}d^2_k,~~k=p,n,
\nonumber\\
\Omega^{\dag}_{\beta}&=&\Omega^{\dag}_p+\Omega^{\dag}_n-2\Omega^{\dag}_{pn},~~
\Omega^{\dag}_{pn}=(b^{\dag}_pb^{\dag}_n)_0-\sqrt{\frac{1}{5}}d^2_p,
\nonumber\\
\hat{N}_{pn}&=&\sum_{m}b^{\dag}_{pm}b_{nm},~\hat{N}_{np}=(\hat{N}_{pn})^{\dag},~~
\hat{N}_k=\sum_{m}b^{\dag}_{km}b_{km},~k=p,n.
\label{omegagen}
\end{eqnarray}
Note that a priory we cannot select one of the two sets of states
$\phi^{(\gamma)}_{JM}$ and $\tilde{\phi}^{(\gamma)}_{JM}$ for gamma band, although  one is symmetric and the other asymmetric against the proton-neutron permutation.
The same is true for the two dipole states isovector candidates.
In \cite{Rad3}, results obtained by using as alternatives a symmetric structure and an asymmetric structure
for the gamma band states were presented. Therein it was shown that the asymmetric structure
for the gamma band does not conflict any of the available data. In contrast, on
considering for the gamma states an asymmetric structure and fitting the model
Hamiltonian coefficients in the manner described  in \cite{Rad2}, for some nuclei a better
description for the beta band energies is obtained. Moreover, in that situation
the description of the E2 transition becomes technically very simple. The results obtained in \cite{Rad2,Rad3} for $^{156}$Gd are relevant in this respect.
For these reasons, here we adopt the option of a proton-neutron asymmetric
gamma band. All calculations performed so far considered equal deformations for protons and neutrons. The deformation parameter for the composite system is:
\begin{equation}
\rho=\sqrt{2}d_p=\sqrt{2}d_n \equiv \sqrt{2}d.
\end{equation}
The factors $N_{J}^{(k)}$ with $k=g,\beta,\gamma,\tilde{\gamma},1,\tilde{1}$ involved in the wave functions are normalization constants calculated in terms of some overlap integrals.

We seek now an effective Hamiltonian for which the projected states (\ref{figcsm}) are, at least to a good approximation, eigenstates in the restricted collective space.
The simplest Hamiltonian fulfilling this condition is:
\begin{eqnarray}
H_{GCSM}&=&A_1(\hat{N}_p+\hat{N}_n)+A_2(\hat{N}_{pn}+\hat{N}_{np})+
\frac{\sqrt{5}}{2}(A_1+A_2)(\Omega^{\dag}_{pn}+\Omega_{np})
\nonumber\\
&&+A_3(\Omega^{\dag}_p\Omega_n+\Omega^{\dag}_n\Omega_p-2\Omega^{\dag}_{pn}
\Omega_{np})+A_4\hat{J}^2,
\label{HGCSM}
\end{eqnarray}
with ${\hat J}$ denoting the proton and neutron total angular momentum.
The Hamiltonian given by Eq.(\ref{HGCSM}) has  only one off-diagonal matrix element in the basis (\ref{figcsm}); that is $\langle\phi^{\beta}_{JM}|H|\phi^{(\tilde{\gamma})}_{JM}\rangle$.
However, our calculations show that this affects the energies of $\beta$ and $\tilde{\gamma}$ bands at the level of a few keV. Therefore, the excitation energies of the six bands are to a  good approximation, given by the diagonal elements:
\begin{equation}
E^{(k)}_J=\langle\phi^{(k)}_{JM}|H|\phi^{(k)}_{JM}\rangle-
\langle\phi^{(g)}_{00}|H|\phi^{(g)}_{00}\rangle,\;\;k=g,\beta,\gamma,1,\tilde{\gamma},\tilde{1}.
\label{EkJ}
\end{equation}
F spin properties of the model Hamiltonian and analytical behavior of energies and wave functions in the extreme limits of vibrational and rotational regimes have been studied 
\cite{Rad2,Rad3,Rad4,Lima,4,Iud}.  Results for the asymptotic regime of deformation suggests that the proposed model generalizes both the two-rotor \cite{LoIu1} and the two-drop models \cite{Grei}. 

Note that $H_{GCSM}$ is invariant under any p-n permutation and therefore its eigenfunctions have a definite parity. We chose one or the other parity for the gamma band, depending on the quality of the overall agreement with the data. We don't exclude the situation when the fitting procedure selects the symmetric $\gamma $ band as the optimal one.
The possibility of having two distinct phases for the collective motion in the gamma band has been considered also in \cite{Novos} within a different formalism.

\section{Extension to a particle-core system}
\label{level3}
\renewcommand{\theequation}{3.\arabic{equation}}
\setcounter{equation}{0}
The particle-core interacting system is described by the following Hamiltonian:
\begin{eqnarray}
H&=&H_{GCSM}+\sum_{\alpha}\epsilon_{a}c^{\dag}_{\alpha}c_{\alpha}-\frac{G}{4}P^{\dag}P
\nonumber\\
&-&\sum_{\tau =p,n}X^{(\tau)}_{pc}\sum_{m}q_{2m}\left(b^{\dag}_{\tau,-m}+(-)^mb_{\tau m}\right)(-)^m -X_{sS}\vec{J}_F\cdot\vec{J}_c, 
\label{modelH}
\end{eqnarray}
with the following notation for the particle quadrupole operator:
\begin{equation}
q_{2m}=\sum_{a,b}Q_{a,b}\left(c^{\dag}_{j_a}c_{j_b}\right)_{2m},~~
Q_{a,b}=\frac{\hat{j}_{a}}{\hat{2}}\langle j_{a}||r^2Y_2||j_b\rangle .
\end{equation}
The core is described by $H_{GCSM}$, while the particle system is described by  the next two terms, standing for a spherical shell model mean field and pairing interactions of the like nucleons, respectively. 
The notation $|\alpha\rangle =|nljm\rangle =|a,m\rangle$ is used for the spherical shell model states.
The last two terms, denoted hereafter as $H_{pc}$, express the interaction between the satellite particles and the core through a quadrupole-quadrupole $qQ$ and a spin-spin force $sS$, respectively. The angular momenta carried by the core and particles are denoted by $\bf{J}_c (= \bf{J}_{p}+\bf{J}_{n})$ and $\bf{J}_F$, respectively. 

The mean field plus the pairing term is quasi-diagonalized by means of the Bogoliubov-Valatin transformation.
The free quasiparticle term is $\sum_{\alpha}E_{a}a^{\dag}_{\alpha}a_{\alpha}$, while the qQ interaction preserves  the above mentioned form, with the factor $q_{2m}$ changed to:
\begin{eqnarray}
q_{2m}&=& \eta^{(-)}_{ab}\left(a^{\dag}_{j_a}a_{j_b}\right)_{2m}+\xi^{(+)}_{ab}\left((a^{\dag}_{j_a}a^{\dag}_{j_b})_{2m}-(a_{j_a}a_{j_b})_{2m}\right),\;\; \rm{where}
\nonumber\\
\eta^{(-)}_{ab}&=&\frac{1}{2}Q_{ab}\left(U_aU_b-V_aV_b\right),\;\;
\xi^{(+)}_{ab}=\frac{1}{2}Q_{ab}\left(U_aV_b+V_aU_b\right).
\end{eqnarray}
The notation $a^{\dagger}_{jm}$ ($a_{jm}$) is used for the quasiparticle creation (annihilation) operator.
We restrict the single particle space to a proton single-j state where two particles are placed.
For the space of the particle-core states, therefore,  we consider the basis defined by:
\begin{equation}
|BCS\rangle\otimes|1;JM\rangle ,
\Psi^{(2qp;J1)}_{JI;M}=N^{(2qp;J1)}_{JI}\sum_{J'}C^{J\;J'\;I}_{J\;1\;J+1}\left(N^{(1)}_{J'}\right)^{-1}\left[(a^{\dag}_ja^{\dag}_j)_J|BCS\rangle\otimes|1;J'\rangle \right]_{IM},
\label{basis}
\end{equation}
where $|BCS\rangle$ denotes the quasiparticle vacuum, while $N_{JI}^{(2qp;J1)}$ is the norm of the projected state. 

\section{Numerical application}
\renewcommand{\theequation}{4.\arabic{equation}}
\setcounter{equation}{0}
\label{level4}
The formalism described above was applied for two isotopes $^{188,190}$Os.  In choosing these isotopes, we had in mind their triaxial shape behavior reflected by the signature 
\begin{equation}
E_{2^+_g}+E_{2^+_{\gamma}}=E_{3^{+}_{\gamma}}.
\end{equation}
Indeed, this equation is obeyed with a deviation of 2 keV for $^{188}$Os and 11 keV for $^{190}$Os.

\subsection{Energies}
We calculated first the excitation energies for the bands described by the angular momentum projected functions
$|g;JM\rangle\otimes|BCS\rangle,~ |\beta;JM\rangle \otimes|BCS\rangle, |\gamma;JM\rangle \otimes|BCS\rangle, |1;JM\rangle \otimes|BCS\rangle, |{\bar 1};JM\rangle\otimes|BCS\rangle $  and the particle-core Hamiltonian $H$. Several parameters, like the structure coefficients defining  the model Hamiltonian and the deformation parameters $\rho$, are to be fixed.  For a given $\rho$ we determined the parameters involved in $H_{GCSM}$ by fitting the excitation energies in the  ground, $\beta$ and $\gamma$ bands, through a least square procedure. We then varied  $\rho$ and kept the value which provides the minimal root mean square of the resulting deviations from the corresponding experimental data. The excitation energies of the phenomenological magnetic bands  are free of any adjusting parameters. In fixing the strengths of  the pairing and the $q.Q$ interactions, we were guided by \cite{Lima}, where spectra of some Pt even-even isotopes  where interpreted  with a particle-core Hamiltonian, the core being described usung the CSM. The two-quasiparticle energy for the proton orbital $h_{11/2}$  was taken as 1.947 MeV for $^{188}$Os and 2.110 MeV for $^{190}$Os , these values being close to the ones yielded by a BCS treatment in the extended space of single particle states.
The parameters mentioned above have the values listed in Table I.
\begin{table}[h!]
\scriptsize{
\begin{tabular}{|cccccccccccc|}
\hline
   &$\rho=d\sqrt{2}$  &   $A_1$[keV]   &     $A_2$[keV]    &   $A_3$[keV]    &   $A_4$[keV] &  $X'_{pc}$[keV]  &  $X_{sS}$[keV] &  $g_p$[$\mu_N$]  &  $g_n$[$\mu_N$] &  $g_F$[$\mu_N$] &r.m.s.[keV]
\\
\hline
$^{188}$Os&  2.2   &   438.7   & -93.8   &   -70.5  &   9.1  &  1.02  &  3.0 &0.828 &-0.028&1.289&16.93\\
$^{190}$Os & 2.0   & 366.1     & 92.6  & 24.0    &    12.2& 1.66& 2.0& 0.7915 & 0.0086 &1.289& 18.63\\  
\hline
\end{tabular}}
\caption{\scriptsize {The structure coefficients of the model Hamiltonian were determined  by a least square procedure. On the last column the r.m.s. values characterizing the deviation of the calculated and experimental energies are also given. The deformation parameter $\rho$ is adimensional. The parameter $X'_{pc}$ is related to $X_{pc}$ by:
$X'_{pc}=6.5\eta^{(-)}_{\frac{11}{2}\frac{11}{2}}X_{pc}.$ }}
\end{table}
Excitation energies calculated with these parameters are compared with the corresponding experimental data, in Figs. 1 and 2. One notes a good agreement of results with the corresponding experimental data.
\begin{figure}[h]
\includegraphics[width=7.5cm]{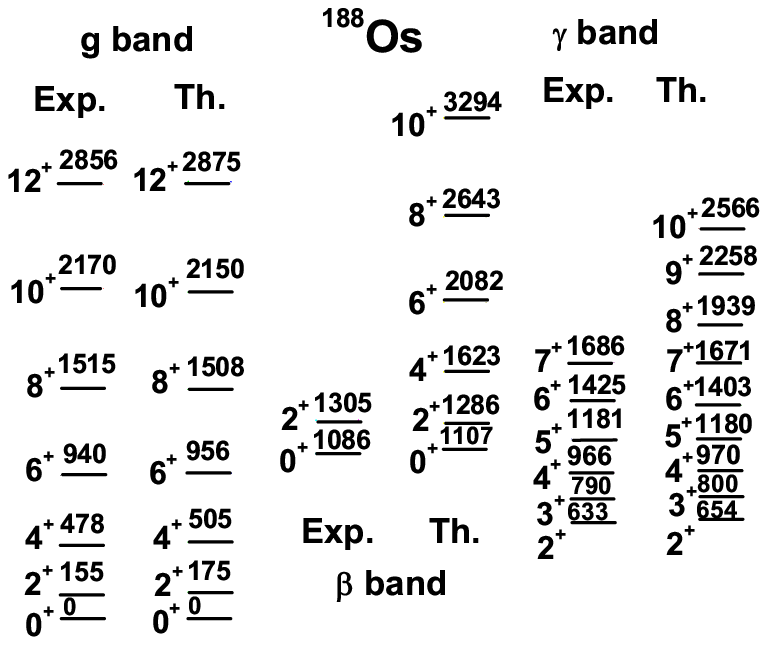}   \includegraphics[width=7.5cm]{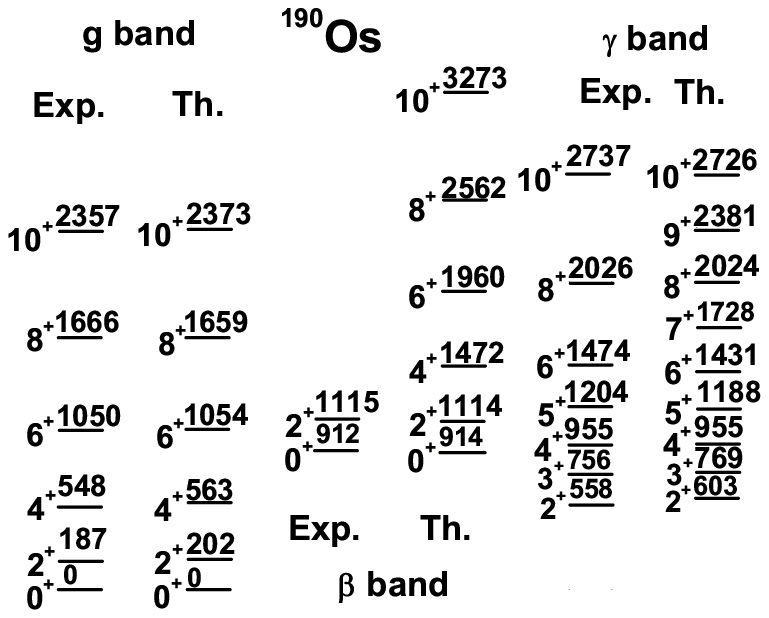}
\begin{minipage}[h]{7cm}
\caption{\scriptsize{Experimental (Exp.) and calculated (Th.) excitation energies in ground, $\beta$ and $\gamma$ bands of $^{188}$Os. Data are taken from \cite{Bal1}. }}
\end{minipage}\ \  
\begin{minipage}[h]{7.5cm}
\caption{\scriptsize{The same as in Fig.1 but for $^{190}$Os with data from Ref.\cite{Bal2}.}}   
\end{minipage}
\end{figure}
\begin{figure}[h!]
\includegraphics[width=7.5cm]{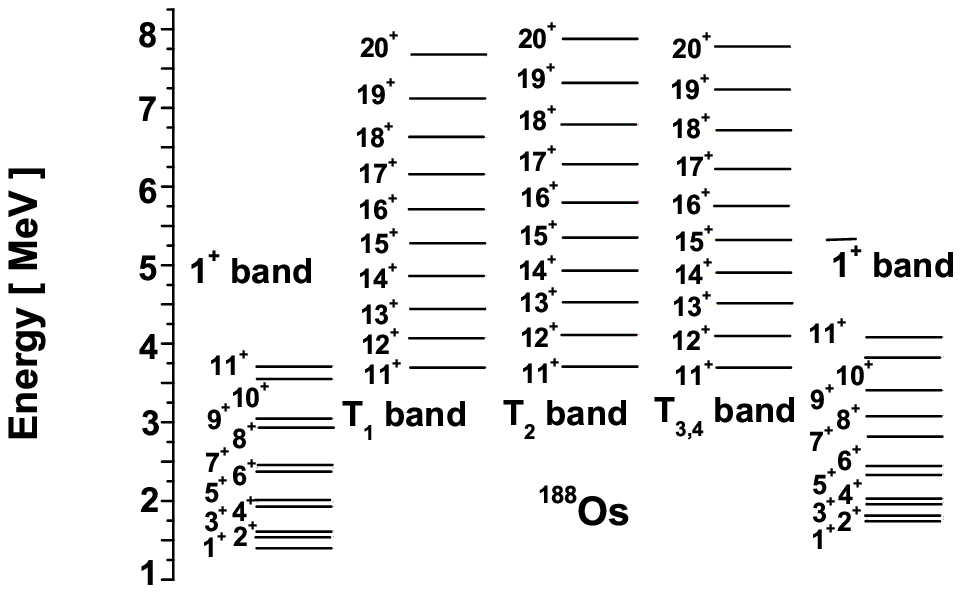} \includegraphics[width=7.5cm]{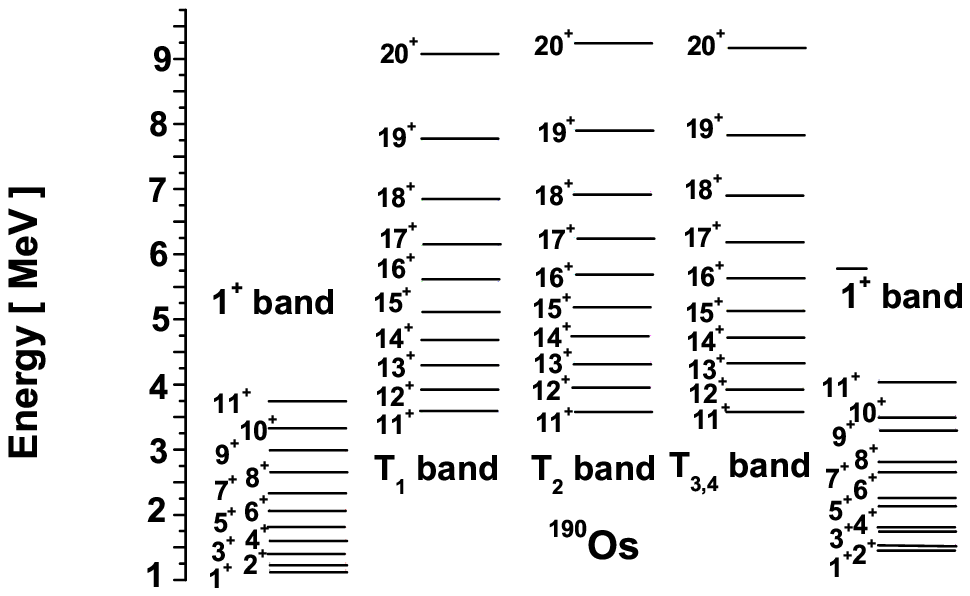}
\begin{minipage}[h]{7.5cm}
\caption{\scriptsize{Excitation energies for the yrast (lower-left) and non-yrast (lower-right) boson dipole states of $^{188}$Os. The twin bands $T_1$ and $T_2$ are also shown.}}    
\end{minipage}\ \
\begin{minipage}[h]{7.5cm}
\caption{\scriptsize{The same as in Fig. 3 but for $^{190}$Os. Here the dipole bands from the lower columns are described by $|1;JM\rangle $ and $|{\bar 1};JM\rangle $.}}
\end{minipage}
\end{figure}
\begin{figure}[h!]
\includegraphics[width=7.5cm]{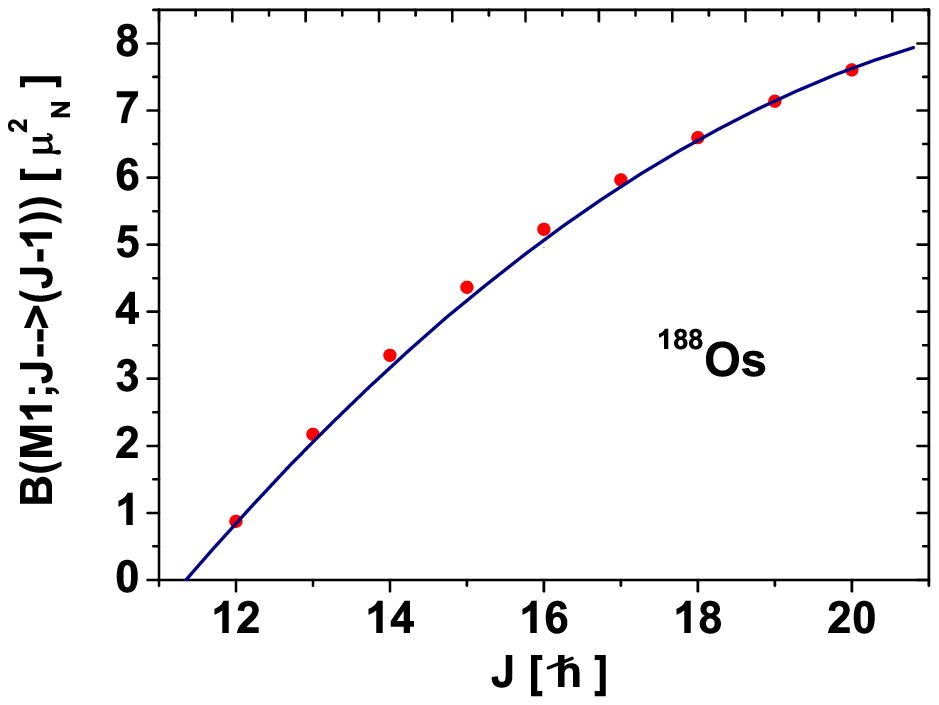} \includegraphics[width=7.5cm]{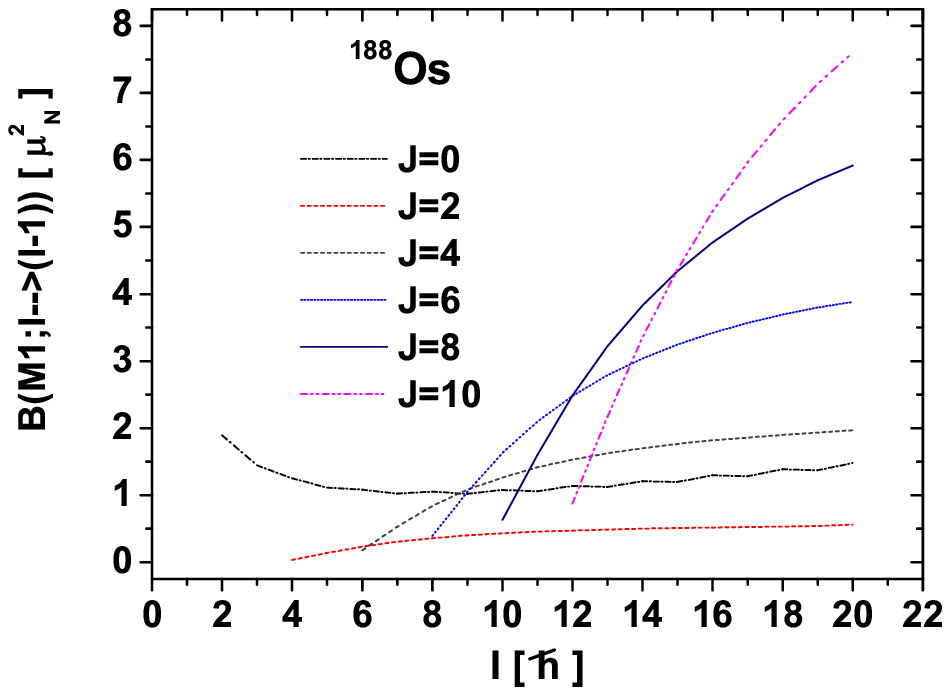}
\begin{minipage}[h]{7.5cm}
\caption{\scriptsize{The BM1 values associated with the dipole magnetic transitions between two consecutive levels in the $T_1$ band of $^{188}$Os. The results are interpolated with a second rank polynomial (full curve). The gyromagnetic factors employed are
$g_p=0.828 \mu_N$, $g_n=-0.028\mu_N$ and $g_F=1,289\mu_N$.}}
\end{minipage}\ \
\begin{minipage}{7.5cm}
\caption{\scriptsize{The magnetic dipole reduced probabilities within the two quasiparticle-core bands  corresponding to the quasiparticle total angular momentum J. The gyromagnetic factors are the same as those used in Fig. 5.}} 
\end{minipage}
\end{figure}
 Unfortunately, there is no data values available concerning the magnetic states. However, in 
\cite{Bal1,Bal2} the 1304.82 keV and 1115.5 keV states in $^{188}$Os and $^{190}$Os respectively, perform an M1 decay to the ground band states. These states could tentatively be associated with the heading states of the two dipole bands which are located at  1400 and 1538 keV, respectively. For $^{188}$Os, the states $|1;JM\rangle$ are not in a natural order from $J\ge 6$. Indeed,
the yrast states belong to the $1^+$ band except the states with $J=6, 8, 10$ which are of ${\bar 1}^+$ type. Similarly, non-yrast states have a ${\bar 1}^+$ character except the states of
$J=6, 8, 10$, which are of $1^+$ type.

\begin{table}
\begin{tabular}{|ccc|ccc|}
\hline
\multicolumn{3}{|c|}{$^{188}$Os}& \multicolumn{3}{c|}{$^{190}$Os}\\
\hline
T$_1$     &T$_2$   &T$_3$=T$_4$    & T$_1$      &T$_2$      &T$_3$=T$_4$\\
\hline
   3.698  & 3.716  & 3.707         & 3.596      & 3.581     & 3.589      \\
   4.081  & 4.117  & 4.099         & 3.962      & 3.931     & 3.947       \\
   4.468  & 4.523  & 4.495         & 4.346      & 4.298     & 4.322      \\
   4.864  & 4.941  & 4.903         & 4.754      & 4.689     & 4.721    \\ 
   5.276  & 5.373  & 5.324         & 5.196      & 5.112     & 5.154    \\
   5.704  & 5.823  & 5.764         & 5.687      & 5.586     & 5.637    \\
   6.152  & 6.295  & 6.224         & 6.259      & 6.139     & 6.199    \\
   6.626  & 6.793  & 6.709         & 6.968      & 6.829     & 6.898    \\ 
   7.128  & 7.320  & 7.224         & 7.916      & 7.757     & 7.836    \\
   7.662  & 7.880  & 7.771         & 9.258      & 9.078     & 9.168     \\
\hline
\end{tabular}
\caption{Excitation energies of the chiral bands $T_1, T_2, T_3=T_4$ given in units of MeV}
\end{table}

If in the expression of H (\ref{modelH}) one ignores the spin-spin term, the resulting Hamiltonian exhibits a chiral symmetry. A chiral transformation in the angular momentum space consists in changing the orientation of one of the axes. Thus the chiral transformation transforms a right oriented trihedral form into a left oriented trihedral form and vice versa. Clearly the spin-spin interaction  breaks the chiral symmetry i.e. this term is not invariant under any chiral transformation. Indeed, changing alternatively the signs of ${\bf J}_F, {\bf J}_p, {\bf J}_n$ one obtains three distinct interactions which, moreover, are different from the initial one. Associating with each of these interactions a band (for more details see Section 5), one obtains a set of four bands which will be conventionally called chiral bands. In figures 3 and 4 the chiral bands $T_1$ and $T_2$ are associated with the actual Hamiltonian given by equation (\ref{modelH}) and the one obtained by the chiral transformation  ${\bf J}_F\to -{\bf J}_F$, respectively, while the bands $T_3$ and $T_4$ are degenerate and correspond to the transformations ${\bf J}_p\to -{\bf J}_p$ and ${\bf J}_n\to -{\bf J}_n$ respectively, applied to the initial reference frame symbolized by $T_1$. The degeneracy is caused by the fact that in both cases the transformed spin-spin interaction is asymmetric with respect to the p-n permutation and therefore their averages with the two quasiparticle-dipole-core states, which are asymmetric, are vanishing. It is remarkable the fact that upon enlarging the particle-core space with the  
$[(2qp)_{J}\otimes \Phi^{(g)}_{J'}]_{IM}$ states, the interaction between the opposite parity $2qp\otimes core$ states, due to the spin-spin term, would determine another two bands of mixed symmetry, characterized also by large M1 rates. The description of such bands will be presented elsewhere. In conclusion, the degeneracy mentioned above is removed  if the space is enlarged such that the parity mixing symmetries are possible. The energies shown in figures 3 and 4 are listed in Table 2.

\subsection{Magnetic properties}

In what follows we give a few details about the calculation of the M1 transition rate.
The magnetic dipole transition operator is defined as:
\begin{equation}
M_{1,m}=\sqrt{\frac{3}{4\pi}}\left(g_pJ_{p,m}+g_nJ_{n,m}+g_FJ_{F,m}\right).
\end{equation}
Considering for the core's magnetic moment the classical definition, one obtains an analytical expression involving the quadrupole coordinates and their time derivatives of first order, which can be further calculated by means of the Heisenberg equation \cite{Rad2,Rad3,Rad4}. Finally, writing the result in terms of quadrupole boson operators and identifying the factors multiplying the proton and neutron angular momenta with the gyromagnetic factors of proton and neutrons, one obtains \cite{Rad3}
\begin{equation}
\left(\begin{matrix}g_p\cr g_n\end{matrix}\right)=\frac{3ZR_0^2}{8\pi k_p^2}\frac{Mc^2}{(\hbar c)^2}\left(\begin{matrix}A_1+6A_4\cr \frac{1}{5}A_3\end{matrix}\right),
\label{gpgn}
\end{equation}
where $Z$ and $R_0$ denote the nuclear charge and radius, while $M$ and $c$ are the proton mass and the velocity of light. $k_p$ is a parameter defining the canonical transformation relating the coordinate and conjugate momenta with the quadrupole bosons, while $A_1, A_3, A_4$ are the structure coefficients involved in $H_{GCSM}$. Within the GCSM the core gyromagnetic factor is
\cite{Rad2}
\begin{equation}
g_c=\frac{1}{2}(g_p+g_n),
\end{equation}
and moreover that might be identified with the liquid drop value, $Z/A$; consequently the canonicity coefficient acquires the expression:
\begin{equation}
k_p^2=\frac{3}{16\pi}AR_0^2\frac{Mc^2}{(\hbar c)^2}\left(A_1+6A_4+\frac{1}{5}A_3\right).
\end{equation}
Inserting this in Eq.(\ref{gpgn}), the gyromagnetic factors are readily obtained. Their values are listed in Table 1.
The fermion gyromagnetic factor corresponds to the proton orbital h$_{11/2}$ with the spin composing term quenched by a factor 0.75.

 With this expression for the transition operator, we also calculated the B(M1) value for the transitions $1^+\to 0^+_g$ and $1^+\to 2^+_g$. The results are 0.2772$\mu^2_N$, 0.0139 $\mu^2_N$ for $^{188}$Os and 0.1752$\mu^2_N$, 0.0085$\mu^2_N$ for
$^{190}$Os. Actually this is consistent with the fact that the nuclear deformations of the considered nuclei are small, which results in there being a relatively small M1 strength for the dipole state $1^+$.
\begin{figure}[h!]
\begin{center}
\includegraphics[width=0.5\textwidth]{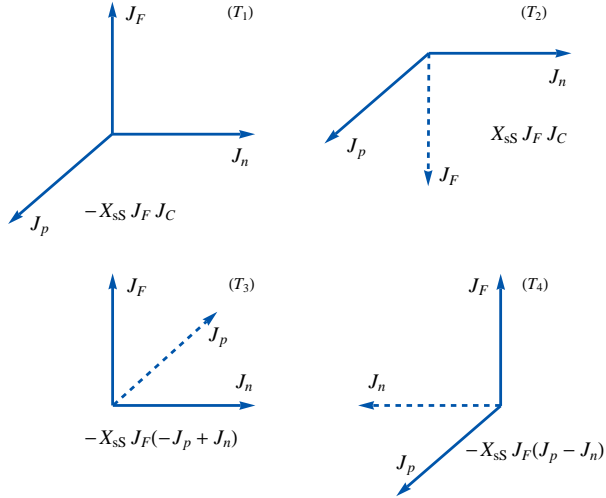}
\end{center}
\caption{ The four frames are related by a chiral transformation. The spin-spin interaction corresponding to each trihedral is also mentioned. They generate the bands $T_{i}$ with i=1,2,3,4, respectively.}
\label{chiralframes}
\end{figure} 
The model used in the present paper was formulated in a previous publication \cite{AAR2014} and applied  for the case of $^{192}$Pt. As we already mentioned before, here we used the same ingredients but for another two triaxial isotopes, $^{188,190}Os$. While in Ref.\cite{AAR2014}  the gyromagnetic factor of neutrons was taken to be $\frac{1}{5}g_p$, here the two factors  are calculated in a self-consistent manner and thus they depend explicitly on the structure coefficients involved in the collective Hamiltonian. Our work proves that the mechanism for chiral symmetry breaking, which also favors a large transverse component for the dipole magnetic transition operator \cite{Frau}, is not unique.
 
The bands $T_1, T_2$ and $T_{3,4}$ , defined above have, do indeed have properties which are specific to the chiral bands: 

i) First of all, as proved in  \cite{AAR2014}, the  trihedral form 
$({\bf J}_p, {\bf J}_n, {\bf J}_F)$ is orthogonal for some total angular momenta of the 2qp-core states, at the beginning of the bands, and almost orthogonal for the next states, and on increasing the total angular momentum the angle is decreased due to the alignment effect caused by rotation. Since the  proton state involved is $h_{11/2}$ and the fermion angular momentum is $J=10$ with the projection $M=J$, this is aligned to the Oz axis, which is perpendicular onto the plane of the orthogonal vectors ${\bf J}_p$ and ${\bf J}_n$;

ii) The energy spacings in the two bands have similar behaviors as function of the total angular momentum. They vary slowly with angular momentum. From Table 2 one notes that for $^{190}$Os
the energy spacing increases with the angular momentum faster than in the case of $^{188}$Os. The reason for this difference is provided by the strength of the qQ-interaction; 

iii) The staggering function $(E(J)-E(J-1))/2J$ is almost constant;

iv) The most significant property is that the B(M1) values for the transition between two consecutive levels are large. The B(M1) values associated to the intra-band transitions are large despite the fact that the deformation is typical for a transitional spherically deformed region; this property is shown in Fig. 5. The fact that the large transition matrix elements are associated with a chiral configuration of the angular momenta involved is illustrated in Fig 6, where one sees that large B(M1) values are achieved for large quasiparticle total angular momentum projection on the symmetry axis. According to figure 5, the M1 strength for the intra-band transitions depends quadratically on the angular momentum of the decaying state. This feature is to be compared with the property of the scissors mode that the the total M1 strength is proportional to the nuclear deformation squared. 

\subsection{More about symmetries}
\renewcommand{\theequation}{5.\arabic{equation}}
\setcounter{equation}{0}
\label{level5}
Our description is different from the ones from the literature in the following respects. The previous formalisms were focussed mainly on the odd-odd nuclei, although a few publications refer also to even-odd \cite{Mukhop} and even-even isotopes \cite{Luo}. Our approach concerns the even-even systems and is based on a new concept.
While until now there have been only two magnetic bands related by a chiral transformation, here we found four magnetic bands with this property, two of them being degenerate.

Indeed, consider the trihedral formas $(\bf{J}_p,\bf{J}_n, \bf{J}_F)$, $(\bf{J}_p,\bf{J}_n, -\bf{J}_F)$, $(-\bf{J}_p,\bf{J}_n, \bf{J}_F)$, $(\bf{J}_p,-\bf{J}_n, \bf{J}_F)$ denoted by the same letters as the associated bands, i.e. $T_1$, $T_2$, $T_3$ and $T_4$, respectively.  To these trihedral forms, four distinct spin-spin interaction terms correspond: $(\bf {J}_F \cdot \bf {J}_c); (-\bf {J}_F \cdot \bf {J}_c); (\bf {J}_F \cdot
(-\bf {J}_p+\bf{J}_n); (\bf {J}_F \cdot(\bf {J}_p-\bf {J}_n)$, each of them affecting the chiral symmetric and degenerate spectrum in a specific way.
Concretely, let us denote by $C_k$ with $k=p,n,F$ the chiral transformation corresponding to the "k" axis and define
\begin{equation}
_{k}\Psi^{(2qp;J1)}_{JI;M}=C_k\Psi^{(2qp;J1)}_{JI;M}, k=p,n,F.
\end{equation}
The average of the model Hamiltonian with the transformed functions is
\begin{equation}
\langle _{k}\Psi^{(2qp;J1)}_{JI;M}| H |_{k}\Psi^{(2qp;J1)}_{JI;M}\rangle =\langle \Psi^{(2qp;J1)}_{JI;M}| (C_k)^+HC_k |\Psi^{(2qp;J1)}_{JI;M}\rangle .
\end{equation}
This equation proves that the four bands are, indeed, determined by the images of the non-invariant part of H through the transformation $C_k$.
According to this equation, the four chiral bands show up upon adding to the space of 2qp-core states given by equation (3.4) the corresponding chirally transformed states. 

Obviously, the four bands are related by the following equations:
\begin{eqnarray}
T_2&=&C_FT_1,\nonumber\\
T_3&=&C_pT_1=R^{n}_{\pi}T_2,\nonumber\\
T_4&=&C_nT_1=R^{p}_{\pi}T_2=R^{F}_{\pi}T_3.
\end{eqnarray}
with $R^{k}_{\pi}$, k=p, n, F, denoting the rotation in the angular momenta space around the axis $k$ with angle $\pi$.  Therefore, if $T_1$ is a right trihedral form then the trihedral forms $T_2, T_3$ and $T_4$ have a left character. Due to this, one may expect that the bands $T_k$ with k=2,3,4, are identical since they have the same chiral nature. This is however not true in our model since the transformations $C_p$ and $C_n$ break not only the chiral symmetry but also the proton-neutron (pn) permutation symmetry. Due to this feature
the bands $T_3$, $T_4$ and $T_2$ are different. Moreover, since for the frames $T_3$ and $ T_4$ the sS term is asymmetric under the  (pn) permutation and consequently its average with wave functions of definite pn parity is vanishing, the corresponding bands are degenerate. Note that upon enlarging the particle-core space with the  $2qp\otimes \Phi^{(g)}_J$ states, the interaction between the opposite parity $2qp\otimes core$ states due to the spin-spin term will determine another two bands of mixed symmetry, characterized also by large M1 rates. In conclusion, the degeneracy mentioned above is removed  if the space is enlarged such that the  parity mixing  is possible. The description of such bands will be presented elsewhere.
The bands of different chiral kinds are conventionally called partner bands. In this respect, the pairs of bands $(T_1,T_2)$, $(T_1,T_3)$, $(T_1,T_4)$ are chiral partner bands.
According to the above equations, the reference frames of similar chiral nature are related by a rotation of angle equal to $\pi$. 

Now let us see how the transformations defined above affect the sS
interaction term. To this end, it is useful to introduce the notation $V_k$ for the interactions specified in figure \ref{chiralframes} which corresponds to the reference frame $T_k$.
Obviously, the following relations hold:
\begin{eqnarray}
V_2&=&C_FV_1,\nonumber\\
V_3&=&C_pV_1=C_nC_FV_1=R^{n}_{\pi}V_2,\nonumber\\
V_4&=&C_nV_1=C_pC_FV_1=R^{p}_{\pi}V_2=R^{F}_{\pi}V_3,\nonumber\\
V_4&=&C_nV_1=C_nC_pV_3=C_FV_3.
\end{eqnarray}
From here, the connections of different chiral transformations  result:
\begin{eqnarray}
C_p&=&C_nC_F,\nonumber\\
C_n&=&C_pC_F,\nonumber\\
C_F&=&C_nC_p.
\end{eqnarray}
Consequently, under the given conditions the set of $C_p, C_n, C_F$ and the unity transformation $I$ form a group. At a superficial glance, this seems to be in conflict with the fact that a chiral transformation changes chirality while the product of two transformations preserves chirality.  We mention however that the equations from above were derived taking into account that
each interaction $V_k$ is invariant under the parity transformation $P$, which simultaneously changes the orientation of the three axes. Due to this result we may extend the notion of the chiral partner bands to any pair of bands $(T_2,T_3)$, $(T_2,T_4)$, $(T_3,T_4)$.

The bands $T_1$ and $T_2$ have different chiralities and thereby they characterize  different nuclear phases. Varying  the interaction strength $X_{sS}$ smoothly from positive to negative values, one may achieve a transition between the two phases. The critical value of the strength is $X_{sS}=0.$ Recall that the  degenerate bands $T_3$ and $T_4$ correspond to this value. On the other hand it has been proved that, generally, the critical point of a phase transition corresponds to a new symmetry \cite{GRC98,Iache2}. Tentatively, the degeneracy of the $T_3$ and $T4$ bands might be related with the symmetry corresponding to the critical spin-spin interaction.

Note that in the absence of the sS interaction, the Hamiltonian is invariant under chiral transformations, and therefore states with left are degenerate with those of right chirality. The model Hamiltonian is also invariant with respect to the pn permutation and consequently its eigenstates are either even or odd with respect to this symmetry. When the sS interaction is switched on both symmetries - chiral and pn permutation - are broken. Thus the energies of left oriented frame are different of those of right character. Moreover, for $T_1$ and $T_2$ the states are of definite pn permutation parity, while upon enlarging the model space, $T_3$ and $T_4$ are split apart and the states become mixture of components of different parities.

Note that fixing the angular momentum orientation may define a certain intrinsic frame while apparently the Hamiltonian is considered in the laboratory frame, due to its scalar character. This is actually not the case. Indeed, the Hamiltonian is invariant under the rotations defined by the components of the total angular momentum but not under those defined by the components of ${\bf J_F}$, ${\bf J_p}$ or ${\bf J_n}$. The pure boson term should be discussed in the framework of the coherent states. Indeed, we recall that the projected states depend on the deformation parameter which implies an asymmetric structure in the intrinsic coordinates. Indeed the projected state is a linear combination of different K components among which one is dominant. Therefore the states are approximatively K oriented and by this the Hamiltonian is considered in a subsidiary intrinsic frame when an angular momentum projected states are used. 

Usually the particle-core formalisms are confronted with the Pauli principle violation. This feature is not encoutered in the present approach. Indeed, the two quasiparticles which are coupled to the core have a maximal angular momentum. On the other hand in a spurious component, which might be generated by the particle number operator, the quasiparticle angular momenta are anti-aligned which results in having a vanishing total angular momentum. We recall the fact that the core is described in terms of phenomenological quadrupole proton and neutron bosons. If we consider a microscopic structure for the aforementioned bosons, then among the composing quadrupole configurations one may find the state of the outer particles. This could be a source for the Pauli principle violation.
Again that does not matter in our case since the 2qp state have a maximal angular momentum while the proton bosons, describing the core, have the angular momentum equal to two.
In conclusion, due to the fact that the 2qp states, which are coupled to the core, are characterized by maximal quantum number K, the Pauli principle is not violated at all, at least as far as    the particle-core interaction is concerned. Of course, since the states of the core are multi-boson states the Pauli principle is violated, which is common to all phenomenological descriptions dealing with bosons. However, if the anharmonic boson Hamiltonian is derived  by means of the Marumori \cite{Marum} boson expansion method, this drawback is certainly removed. Since our description uses phenomenological quadrupole bosons, the core's feature mentioned above does not show up.

In order to stress on the novel features introduced by the present paper, it is worth  summarizing the differences with respect to the approach of \cite{AAR2014}:

a) The nuclei chosen for applications are different; 

b) The M1 properties are studied with different transition operators; 

c) Here we discuss the phase transition between left and right chiral nuclear phases; 

d) The critical point of this transition is characterized by a new symmetry associated with the two degenerate bands $T_3$ and $T_4$; 

e) On enlarging the 2qp-core space by adding states of even parity with respect to the proton-neutron permutation, as for example the states $2qp\otimes \Phi^{(g)}_J$, and then diagonalizing the spin-spin interactions $V_3$ and $V_4$ in the extended space, the degeneracy of the two bands $T_3$ and $T_4$ is lifted. Moreover, with the interactions $V_{3}$ and $V_{4}$ one associates two bands of mixed parity and left chirality. The composing states  have the 2qp-dipole core functions as dominant components. The resulting non-degenerate bands have mixed parity and left chiralities and  define  new $T_3$ and $T_4$ bands, denoted hereafter by $T_{3}^{\prime}$ and $T_{4}^{\prime}$. The band $T_{3}^{\prime}$ is composed of states having the maximum component for the 2qp-dipole core state yielded by diagonalizing $V_{3}$. Similarly one defines the band $T_{4}^{\prime}$ but with the sS interaction $V_{4}$.  Note that while $T^{\prime}_{3}$ and $T^{\prime}_{4}$ bands have strong intra-band M1 transitions the other two bands provided by diagonalization, denoted by $T_{3}^{\prime\prime}$ and $T_{4}^{\prime\prime}$, are connected with $T^{\prime}_{3}$ and 
$T^{\prime}_{4}$ respectively, by weak inter-band M1 transitions. The angular momentum alignment is a slow process and therefore the angle between ${\bf J}_p$ and ${\bf J}_n$ remains large, which contrasts the scissors mode which is characterized by small angle between the symmetry axes of the proton and neutron systems. From this point of view the chiral states are of shears type rather than of scissors type. The intra-band M1 strength depends quadratically on the angular momentum, while the M1 strength for a scissors mode is proportional to the nuclear deformation squared.

\section{Conclusions}
Here we considered two proton quasiparticle bands but alternatively we could chose two neutron quasiparticles  or one proton plus one neutron quasiparticle bands. Of course, the latter bands would give a description of an odd-odd system. We already checked that a two neutron quasiparticle band is  characterized by a non-collective M1 transition rate. This feature suggests that, indeed, the orbital magnetic moment carried by protons plays an important role in determining a chiral magnetic band. The core is described through angular momentum projected states from a proton and neutron coherent state as well as from its lowest order polynomial 
excitations. Among the three chiral angular momentum components two are associated with the core and one with a two quasiparticle state. In contradistinction, the previous descriptions devoted to odd-odd system, use a different picture. The core carries one angular momentum and, moreover, its shape structure determines the orientation of the other two angular momenta associated with the odd proton and odd neutron, respectively. For odd-odd nuclei several groups identified twin bands in medium mass regions \cite{Petrache96,Simon2005,Vaman,Petrache06}  and even for heavy mass regions\cite{Balab,Frau97}.
Theoretical approaches are based mainly on  a triaxial rotor-two quasiparticle coupling, which was earlier formulated and widely used by the group of Faessler \cite{Fas1,Fas2,Fas3,Fas4}.
For a certain value of the total angular momentum, the angular momenta carried by the three components are mutually orthogonal. This picture persists for the next two angular momenta and then increasing the the rotation frequency, the core spins are gradually aligned. Subsequently, the quasiparticle angular momentum is also aligned with the resulting spin. This is the mechanism which develops a $\Delta I$=1 band. The new features of the present approach are underlined by comparing it with the formalism of \cite{AAR2014}.

As mentioned before, the even-even nuclei which might be good candidates for use in exploring the chiral properties are those of a triaxial shape. Moreover, the satellite protons are to be in a shell of large angular momentum. In this way the protons orbital angular momentum provides a consistent contribution to the M1 strength. Also, the chosen nucleus must belong to a transitional spherically deformed region, i.e. it exhibits a small nuclear deformation. In the present approach the chiral states are of 2qp-core dipole state type which implies that the core  has a low lying dipole band. 
The numerical results for the chosen nuclei are consistent with the commonly accepted signatures of the chiral bands. The intra-band M1 strength has a quadratic dependence on the state angular momentum, which contrasts the case for the scissors mode, whose strength is proportional to the  nuclear deformation squared. The chiral bands are characterized by large angles between the proton and neutron symmetry axes, while for the scissor mode this angle is very small. The calculated M1 strength for the transition from a chiral band to the band generated by the $2qp\otimes \Phi^{(g)}_J$ is small, which confirms the fact that the considered states have a different nature than the scissors-like states.

The strength parameters characterizing the core were fixed by fitting some energies from ground, $\beta$ and $\gamma$ bands. The agreement with experimental data in the core bands is very good.
Unfortunately, no data for the dipole bands are available for the chosen nuclei. 
Experimental data for chiral bands in even-even nuclei are desirable. These would encourage us to extend the present description to a systematic study of the chiral features in even-even nuclei.
The present paper has the merit of drawing the attention to the fact that such states, organized in twin bands, exist. We believe that predicting new features of the nuclear system and describing the existing data are equally important ways of achieving progress in the field.

{\bf Acknowledgment.} This work was supported by the Romanian Ministry for Education Research Youth and Sport through the CNCSIS project ID-2/5.10.2011.


\begin{references}

\bibitem{LoIu1} N. Lo Iudice and F. Palumbo, Phys. Rev. Lett. {\bf 41}, 1532 (1978).
\bibitem{LoIu2} G. De Francheschi, F. Palumbo and N. Lo Iudice, Phys. Rev. {\bf C29} (1984) 1496.
\bibitem{LoIu3} N. Lo Iudice, Phys. Part. Nucl. {\bf 25 }, 556, (1997).
\bibitem{Zawischa} D. Zawischa, J. Phys. G{\bf 24}, 683, (1998).
\bibitem{Frau}S. Frauendorf, Rev. Mod. Phys.  {\bf 73} (2001) 463.
\bibitem{Jenkins} A. P. Jenkins et al., Phys. Rev. Lett. {\bf 83} (1999) 500.
\bibitem{AAR2014} A. A. Raduta, C. M. Raduta and A. Faessler, J. Phys. G: Nucl. Part. Phys {\bf 41} (2014) 035105.
\bibitem{Rad2}A. A. Raduta, A. Faessler and V. Ceausescu, Phys. Rev. {\bf C 36} (1987) 439.
\bibitem{Rad1}A. A. Raduta {\it et al.}, Phys. Lett. {\bf B 121}1; Nucl. Phys. {\bf A 381}
(1982) 253.
\bibitem{Sheline} R. K. Sheline, Rev. Mod. Phys. {\bf 32}, 1, (1960); M. Sakai, Nucl. Phys. {\bf A 104}, 301 (1976). 
\bibitem{Rad3}A. A. Raduta, I. I. Ursu and D. S. Delion, Nucl. Phys. {\bf A 475} (1987) 439.
\bibitem{Rad4}A. A. Raduta and D. S. Delion, Nucl. Phys. {\bf A 491} (1989) 24.
\bibitem{Lima}A. A. Raduta, C. Lima and Amand Faessler, Z. Phys. A - Atoms and Nuclei {\bf 313}, (1983), 69.
\bibitem{4} N. Lo Iudice, {\it et al.}, Phys. Lett. {\bf B 300} (1993) 195; Phys. Rev. {\bf C 50} (1994) 127.
\bibitem{Iud}N. Lo Iudice, A. A. Raduta and D. S. Delion, Phys. Rev. {\bf C50} (1994) 127.
\bibitem{Grei} V. Maruhn-Rezwani, W. Greiner and J. A. Maruhn, Phys. Lett. {\bf 57 B} (1975) 109.
\bibitem{Novos} A. Novoselski and I. Talmi, Phys. Lett. {\bf 60 B}, 13 (1985).
\bibitem{Bal1}Balraj Sinh, Nuclear Data Sheets {\bf 95} (2002) 387.
\bibitem{Bal2}Balraj Sinh, Nuclear Data Sheets {\bf 99} (2003) 275.
\bibitem{Mukhop} S. Mukhopadhyay, {\it et al.}, Phys. Rev. Lett. {\bf 99}, 172501 (2007).
\bibitem{Luo} Y. X. Luo, {\it et al.}, Phys. Lett. {\bf B 670}, 307 (2009).
\bibitem{GRC98} A. Gheorghe, A. A. Raduta and V. Ceausescu, Nucl. Phys. {\bf A 637 } (1998) 201.
\bibitem{Iache2} F. Iachello, Phys. Rev. Lett. {\bf 85} (2000) 3580.
\bibitem{Marum} T. Marumori, M. Yamamura, A. Tokunaga and K. Takada, Prog. Theor. Phys. {\bf 32} (1964) 726.
\bibitem{Petrache96}C. M. Petrache {\it et al.,} Nucl. Phys. {\bf A597} (1996) 106.
\bibitem{Simon2005}A. J. Simon {\it et al.,} Jour. Phys. G: Nucl. Part. Phys {\bf 31} (2005) 541.
\bibitem{Vaman}C. Vaman, {\it et al.}, Phys. Rev. Lett. {\bf 92} (2004) 032501.
\bibitem{Petrache06}C. M. Petrache, {\it et al.}, Phys. Rev. Lett. {\bf 96} (2006) 112502.
\bibitem{Balab} D. L. Balabanski, {\it et al.,} Phys. Rev. C {\bf 70} (2004) 044305.
\bibitem{Frau97}S. Frauendorf and J. Meng, Nucl. Phys. {\bf A 617} (1997) 131.
\bibitem{Fas1} H. Toki and Amand Faessler, Nucl. Phys. {\bf A 253} (1975) 231.
\bibitem{Fas2} H. Toki and Amand Faessler, Z. Phys.. {\bf A 276} (1976) 35.
\bibitem{Fas3} H. Toki and Amand Faessler, Phys. Lett. {\bf B 63} (1976) 121.
\bibitem{Fas4} H. Toki, H. L. Yadav and Amand Faessler, Phys. Lett. {\bf B 66} (1977) 310.
\end{references}
\end{document}